\documentclass[letterpaper,twocolumn,10pt]{article}
\usepackage{usenix2019_v3}
\usepackage{url}
\usepackage{tikz}
\usepackage{amsmath}
\usepackage{amssymb}
\usepackage{mathtools}
\usepackage{wrapfig}
\usepackage{bbm}
\usepackage{float}

\usepackage{filecontents}

\usepackage[colorinlistoftodos]{todonotes}
\usepackage{subfigure}

\newtheorem{definition}{Definition}[section]

\newcommand{\Reals}{\mathbb{R}}

\newcommand{\Ex}{\mathbb{E}}

\newcommand{\bW}{\boldsymbol{W}}
\newcommand{\bX}{\boldsymbol{X}}

\newcommand{\bT}{\boldsymbol{T}}
\newcommand{\bM}{\boldsymbol{M}}

\newcommand{\hM}{\widehat{M}}
\newcommand{\hf}{\widehat{f}}

\newcommand{\VoD}{{\text{VoD}}\text{(User  i, Advertiser j)}}
\newcommand{\hVoD}{\widehat{\text{VoD}}\text{(User  i, Advertiser j)}}
\newcommand{\hVoDUser}{\widehat{\text{VoD}}^{\text{norm}}}
\newcommand{\hmupre}{\hat{\mu}_\text{pre}}
\newcommand{\hmupost}{\hat{\mu}_\text{post}}
\newcommand{\hvarpre}{\widehat{\text{var}}_\text{pre}}
\newcommand{\hvarpost}{\widehat{\text{var}}_\text{post}}
\newcommand{\hmudiff}{\hat{\mu}_\text{diff}}
\newcommand{\hvardiff}{\widehat{\text{var}}_\text{diff}}

\begin{document}

\date{}

\title{Zorro: A Model Agnostic System to Price Consumer Data}

\author{
{\rm Anish Agarwal}\\
MIT
\and
{\rm Munther Dahleh}\\
MIT
 \and
{\rm Devavrat Shah}\\
MIT 
 \and
 {\rm Dylan Sleeper}\\
MIT
 \and
 {\rm Andrew Tsai}\\
MIT
 \and
  {\rm Madeline Wong}\\
MIT
 \and
} 

\maketitle

\begin{abstract}
	Personal data is a key ingredient in showing web users targeted ads - the economic backbone of the web. 
Still, there are two major inefficiencies in how such data is bought and sold online: (1) users do not decide what information is released nor do they get paid for this privacy loss; (2) algorithmic advertisers are stuck in inefficient long-term contracts where they purchase user data without knowing how much value it provides. 
This paper proposes a system, Zorro, which aims to rectify the aforementioned two problems. 

As the main contribution, we provide a natural, ``absolute" definition of the ``Value of Data” (VoD) -- for any quantity of interest, it is the delta between an individual's value and the population mean. The challenge remains how to operationalize this definition such that it is independent of a buyer's model for the VoD. We propose a model agnostic solution by relying on matrix estimation, a rapidly growing field within machine learning, and show how it can be used to estimate click-through-rate (CTR), for example. 

Regarding (2), Zorro (and this VoD definition) empowers advertisers to measure value of user data: (i) on a query-by-query basis; (ii) based only on \textit{increase in accuracy} it provides in estimating CTR. This is in contrast with inefficient long-term contracts advertisers are engaged in now with third-party data sellers. We highlight two experimental results on a large real-world ad click dataset. (i) Our CTR estimation system has a $R^2$ of $0.58$, in line with state-of-the-art results for comparable problems (e.g. content recommendation). Crucially, our system is model agnostic, i.e., we estimate CTR for a given user and advertiser \textit{without accessing an advertiser's proprietary models}, a necessary property of any such pricing system. (ii) With respect to our definition of VoD, experiments show selling user data has incremental value in estimating CTR ranging from 30\% to 69\% depending on advertiser category. Roughly, this translates to at least a USD 16 Billion loss in value for advertisers if user data is not provided.

Regarding (1), in addition to allowing users to get paid for data they share, we extend our system design for when users provide explicit intent for types of ads they want to see.

\end{abstract}

\section{Introduction} \label{sec:introduction}
\begin{figure*}
\centering
\subfigure[]{%
\label{fig:ecosystem_overview_1}%
\includegraphics[width=0.45\linewidth,height=2.5cm]{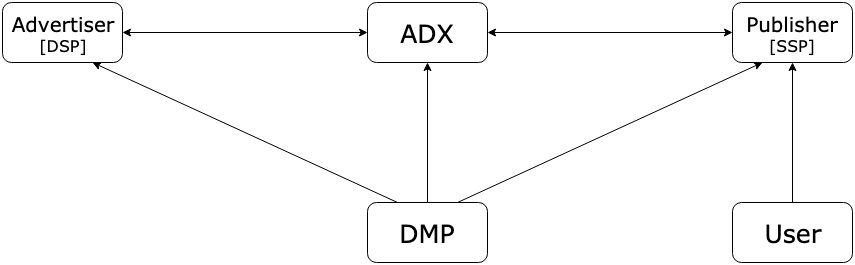}}%
\hfill
\subfigure[]{%
\label{fig:ecosystem_overview_2}%
\includegraphics[width=0.45\linewidth,height=2.5cm]{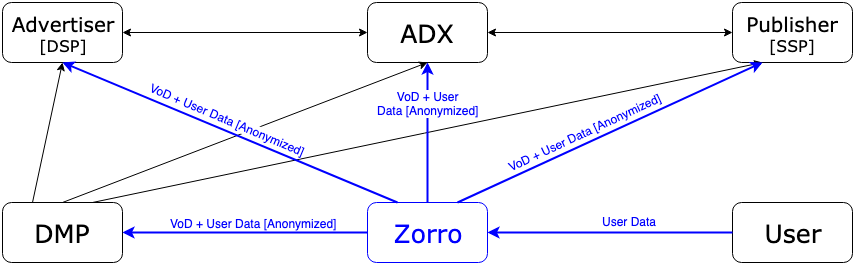}}%
\vspace{-3mm}
\caption{Figure \ref{fig:ecosystem_overview_1} - we depict how user data flows between the various entities in the current ecosystem. Figure \ref{fig:ecosystem_overview_2} - we depict the change to user data flow with Zorro in place. All entities receive (anonymized) user data \textit{and} the estimated VoD.}
\label{fig:RTB_overview}
\end{figure*}

\vspace{-2mm}
\begin{quote} 
\centering 
``\textit{Personal data is the new oil of the internet and the new currency of the digital world }"
\end{quote}
\vspace{-2.4mm}
- Meglena Kuneva, European Consumer Commissioner

\vspace{2mm}

\noindent Over the past two decades, online advertising has become the economic backbone of the web. According to a study by PriceWaterhouseCoopers and the Internet Advertising Bureau (IAB), non-search
\footnote{Non-search includes display ads, audio \& video ads, classifieds and lead generation. For simplicity, we collectively call them display ads (see~\cite{2018iab} for details).}
 online advertising revenue has grown from USD 50 million in 1996 to USD 54 Billion in 2018, with a 23\% year-on-year growth from 2017 to 2018 (cf.~\cite{2018iab}). 
Furthermore, personal data has been one of the key resources advertisers use to show users targeted, relevant display ads.


\vspace{2mm}

\noindent \textbf{Inefficiencies in Current System.} There still exist two major inefficiencies in the way such data is both bought and sold: (1) from a user's perspective, they do not get to decide what information about themselves is released nor do they get paid for this loss of privacy; (2) from an advertiser's perspective, they have to purchase user data without knowing how much value it is going to provide to them. 

These inefficiencies have led to a swath of societal issues including: (i) increasing ad fraud, where 20\% to 30\% of online ad data sold has been deemed to be fraudulent (cf.~\cite{wiredClickFraud2006}); (ii) an increasing consumer backlash for loss of privacy from unsanctioned data sharing such as in the Facebook-Cambridge Analytica case (cf.~\cite{cadwalladr2018revealed}); (iii) an inability to litigate against companies with massive data breaches such as Equifax (cf.~\cite{berghel2017equifax}).

To have a meaningful dialogue about such issues, an important and necessary first step is defining what the ``Value of Data" (VoD) is. However, there does not exist a precise definition of the VoD, which is independent of a buyer's model for VoD. This is why firms are engaged in inflexible long-term data contracts, instead of paying on a query-by-query basis based on the incremental value the data provides.

\vspace{2mm}

\noindent \textbf{Why Data is a Unique Commodity.} The challenge in designing data pricing systems stems from the nature of data as an asset: (i) user data is freely replicable once sold; (ii) advertisers vary widely and correspondingly, so do their models for the VoD; (iii) the usefulness of user data is difficult to verify a priori before running it through a buyer's model, something buyers are not willing to do before purchasing the data. See \cite{2019marketForData} for a more complete overview.

\vspace{2mm}

\noindent \textbf{Our Focus: Click-Through-Rate Estimation.} We focus our efforts on the sale of online user cookie data (e.g. a user's demographic data and browsing history) for display advertising and the prediction task of estimating user click-through-rate (CTR) for a given ad and user, the key quantity of interest in online display advertising. We do so because the sale of online cookie data for this setting is one of the large sources of data exchanged (and revenue generated) in modern web-scale systems and the associated network engineering infrastructure is extremely performant; transactions between display advertisers and third-party data sellers normally occur in less than 100 milliseconds, the time required for a user to load a webpage and for an ad to be served (cf.~\cite{wang2017display}). 

Thus if one can design a feasible solution for this setting, there are many further applications. An important business-to-business example is the sale of potential consumer leads in insurance and mortgage markets using the increasingly adopted Ping-Post Architecture (cf.~\cite{Boberdoo2019}). 

\subsection{Our Contribution}\label{sec:contributions}

We propose a system, Zorro (see Figure \ref{fig:ecosystem_overview_1}-\ref{fig:ecosystem_overview_2}), which aims to address the following aforementioned problems. Our contributions are summarized below.

\vspace{2mm}

\noindent \textbf{VoD Definition.} In Section \ref{sec:system_description}, we give a natural definition of the VoD -- for any quantity of interest (in our case, CTR) it is the delta between an individual's value and the population mean. The key challenge we try to solve through our system design is how to operationalize such a definition in an ``absolute" sense, \textit{independent of a  buyer's models for the VoD}. 

\vspace{2mm}

\noindent \textbf{System Design.} In Section \ref{sec:system_description} we give an overview of the key functionalities/modules of Zorro. (1) An inference system to operationalize our definition of the VoD. This is a challenging inference problem as we only get access to sparse, noisy click data from users. We propose to solve it by reducing CTR estimation to an instance of matrix estimation, where the rows of the induced matrix are users, columns are advertisers and entries are historical click data. Thus we can exploit the rich matrix estimation literature, and utilize data-driven, model agnostic matrix estimation algorithms that have become a workhorse in large-scale content recommendation systems. (2) A module to identify and categorize online ads in an automated fashion. (3) An interface between users, Zorro and advertisers that ensures user data is not inadvertently ``leaked" to advertisers without user permission and ensures advertisers only pay for user data based on the increased in accuracy it provides to them (in particular, to estimate CTR). 

\vspace{2mm}

\noindent \textbf{Experiments.} In Section \ref{sec:experiments}, we perform four set of experiments to verify the implementation of the modules described above. (1) On a large real-world advertising click dataset, we show we accurately estimate the CTR for a given user and advertiser ($R^2$ of $0.58$) independently of an advertiser's model, the crucial property we require to operationalize our definition of the VoD. (2) On the same dataset, we show that with respect to our definition of the VoD, there is large variation in the value of user data depending on the advertiser's category (30\% to 69\% depending on category). (3) We show we can accurately classify (83\% True Positive and 98\% True Negative) and categorize ads (53\% correct classification out of 341 categories) by testing our implementation on popular websites. (4) We show we can indeed prevent user data from being ``leaked" to advertisers by counting the number of repeats ads a user sees depending on whether Zorro is turned ``on" or ``off" (number of repeats drops three-fold when ``on") .

\vspace{2mm}

\noindent \textbf{Forward Looking Intent.} In Section \ref{sec:forward_intent}, we show how one can generalize our model to capture forward-looking intent of users. We do so by extending the fundamental data structure of interest, the matrix of (users $\times$ advertisers), to an order-3 tensor of (user $\times$ advertisers $\times$ user-intent) and furnishing an algorithm that exploits the additional structure when estimating CTR. We verify experimentally that our proposed algorithm does effectively exploit the additional latent structure in the (user $\times$ advertisers $\times$ user-intent) tensor by comparing against an appropriate baseline ($R^2$ of 0.53 vs. 0.17).

\subsection{Related Work}\label{sec:related_work}

We divide related work into the three major buckets: (1) real-time online systems to sell user data; (2) data mining to estimate value of user data; (3) theoretical market mechanisms to sell information.

\vspace{2mm}

\noindent \textbf{Real-Time Systems to Sell User Data.} We compare with two proposed systems (cf. \cite{riederer2011sale, Brave2019}) that aim to design real-time pricing mechanisms of online user data in the context of advertising. The key difference is in Zorro, we propose a precise definition of the VoD. Without such a definition, which is agreed upon by both buyer and seller, price setting remains a challenge. 

In \cite{riederer2011sale}, they propose selling historical user data by posting a price each time a user loads a webpage and running an ``infinite supply" auction, i.e., if an advertiser bids above the posted price, then acquire user data for that webpage load. This posted price is then constantly updated for each subsequent webpage load by that user. Such a mechanism potentially has limitations. (i) Optimal online price setting for data in such systems can be infeasible using standard methods (e.g. a no-regret update mechanism). For example, no-regret update schemes are feasible only if user data is relatively static. However, user's browsing patterns are constantly changing and so it is not immediately clear how to adapt a no-regret update scheme to maximize revenue for such a case. (ii) Advertisers either get all of the data and pay the posted price or receive none of it. This could potentially lead to a large loss in revenue. Zorro aims to circumvent these two problems by dividing the transaction into two segments: (i) an offline segment, where each advertiser negotiates a contract with Zorro on how much they value a \textit{marginal increase in accuracy}; (ii) an online segment, where Zorro estimates the marginal increase in accuracy without access to the advertisers models. Advertisers then pay based on the real-time estimated increase in accuracy and the offline contract (see Section \ref{sec:dsp_zorro_interface} for details of the implementation). Importantly, both the offline and online segments of the contract in Zorro are \textit{entirely auditable} by a third-party leading to a more robust system.

|n \cite{Brave2019}, they propose a blockchain based system and advocate for a new cryptocurrency named Basic Attention Token (BAT), which allows user to get paid in BAT for viewing ads online. For their system to be widely adopted, it would require an overhaul of the entire, very well-established online advertising ecosystem (see Section \ref{sec:online_ad_market}), and need users, advertisers and publishers to all buy into this new cryptocurrency. In contrast, Zorro can be plugged into the current ad ecosystem in a straightforward manner agnostic to a buyer's model. It essentially functions as a novel third-party data seller (i.e. Data Management Platform - see Section \ref{sec:online_ad_market}), where users get paid for providing their data and for advertisers pay based on the estimated increase in accuracy they experience. Further in \cite{Brave2019}, they do not provide details on how pricing of user data is done, a crucial part of any such system.


\vspace{2mm}

\noindent \textbf{Data Mining to Estimate Value of User Data.} Given how important the problem of understanding the role of user data in ad targeting, there has been a lot of empirical work in this field. We highlight a few representative works, \cite{beales2010value, yan2009much, tsai2011effect, gill2013followMoney}. The focus of such work has been to mine large historical datasets to estimate the \textit{average} value of user data. They do indeed find that user data can increase value ranging from 20-60\%, depending on the user and advertiser type, which is line with our findings. However, these works propose highly specific approaches that require intimate knowledge of how a buyer will extract information/value from the data. In contrast, by applying matrix estimation to estimate CTR, we propose a general-purpose model agnostic method that allows advertisers to estimate VoD on a query-by-query basis.

\vspace{2mm}

\noindent \textbf{Theoretical Market Mechanisms to Sell Information.} Some works that propose theoretical market designs to sell information include \cite{2019marketForData, Bergemann2018Information}. We consider our work to be complimentary to theirs as a fundamental quantity that needs to be estimated in such works is the VoD. Zorro proposes a natural, data-driven definition of VoD that can potentially be used to operationalize such systems. Further the works listed have a single buyer and multiple data sellers (e.g. a supply-chain company buying data streams for demand forecasting). This is in contrast with Zorro which is meant for direct, repeated sales of data between multiple buyers and a single seller.

\subsection{Organization of Paper}

The rest of the paper is organized as follows: (1) In Section \ref{sec:online_ad_market}, we give an overview of the online advertising ecosystem that matches advertisers to users and the role of third-party data sellers within it. (2) In Section \ref{sec:system_description}, we precisely define the VoD and justify it. We also give a detailed conceptual overview of all of the key components of the Zorro architecture. (3) In Section \ref{sec:experiments}, we rigorously test the various components of our architecture on large real-world datasets. (4) In Section \ref{sec:forward_intent}, we show how to extend Zorro to include forward-looking intent and corroborate our proposal with experiments. (5) Lastly, in Section \ref{sec:conclusion}, we discuss the role of Zorro in designing a general-purpose data recommendation system.  

\section{Overview of Real-Time Bidding Ecosystem}\label{sec:online_ad_market}

The Real-Time Bidding (RTB) ecosystem refers to the entire engineering infrastructure associated with buying and selling ads in real-time as users load webpages. We give a brief summary of the key players in the RTB ecosystem below. Refer to \cite{wang2017display} for a detailed description of RTB. See Figure \ref{fig:ecosystem_overview_1} for an overview of the current ecosystem.

\vspace{2mm}

\noindent  \textbf{Advertisers and Demand Side Platforms (DSPs).} Advertisers are companies that want to market their product online. Due to the engineering complexity of the RTB ecosystem, advertisers normally go through a DSP to make bids into the RTB system on their behalf. Thus DSPs essentially serve as algorithmic online advertising agencies. 

For our purposes it is not necessary to differentiate between advertisers and DSPs, and we call the collective entity ``advertisers" for simplicity. 

\vspace{2mm}

\noindent  \textbf{Publishers and Supply Side Platforms (SSPs).} Websites that users visit (e.g. Nytimes, ESPN) are called publishers. When a user loads a publisher webpage, the publisher sends a request out into the RTB system to serve the user ads. Analogous to advertisers, publishers go through SSPs to make requests in the RTB system on their behalf. 
	
\vspace{2mm}

\noindent  \textbf{Ad Exchanges (ADX).} ADX serve as the interface between DSPs and SSPs by matching ad requests from SSPs to bids made by DSPs. They do so by using a combination of first-price and second-price auctions.

\vspace{2mm}

\noindent  \textbf{Data Exchanges a.k.a Data Management Platforms (DMPs).} DMPs serve DSPs, SSPs and ADXs historical user data (usually in real-time) to feed into their models to allow for better, more targeted matchings between users and ads.

\vspace{2mm}

\noindent \textbf{Our Focus is Advertiser-DMP Interface.} The majority of demand (and source of revenue) for historical user data is from advertisers. User data is crucial to them as their models require such data to make more accurate bids by evaluating CTR. That is why we focus the discussion of this paper to solve inefficiencies between how Users -- DMPs -- Advertisers interface (see Section \ref{sec:introduction} for an overview of the current market inefficiencies) . However none of our system design is specific to the transaction between DSPs and DMPs, and can be extended to include SSPs and ADXs as well (see Section \ref{sec:system_description}). See Figure \ref{fig:ecosystem_overview_2} for an overview of the proposed change to ecosystem with Zorro.



\section{Zorro System Overview}\label{sec:system_description}

As mentioned in Section \ref{sec:contributions}, we focus our system design on the following functionalities that are required for Zorro to be feasible in practice:
\begin{enumerate}
	\item \textbf{VoD Definition.} We provide a natural, ``absolute" definition of the VoD that captures the incremental value that data provides in estimating CTR. Importantly, our definition is independent of the models of the buyer. 
	\item \textbf{VoD Computation.} To operationalize our definition of the VoD, the key requirement is that for a given advertiser, we need to estimate the value of a user's data independently of an advertiser's models. We show how one can achieve this using matrix estimation to accurately estimate the CTR in a model agnostic manner.
	\item \textbf{Ad Identification \& Categorization.} A necessary step to do CTR estimation, is to accurately identify when users have clicked on an ad. Further to effectively group advertisers, we need to categorize advertisers in a meaningful and automated way. We show how to do both ad identification and categorization by tying together existing open source software packages.
	\item \textbf{User -- Zorro -- Advertiser Interface.}  We propose an interface/dynamic that ensures advertisers pay for user data based only on the increase in accuracy it brings. And correspondingly, ensures user data is not inadvertently ``leaked" to advertisers without them paying for it. 
\end{enumerate}

In Sections \ref{sec:defining_value_of_data} - \ref{sec:dsp_zorro_interface}, we explain how Zorro performs the above four functions respectively. Our focus will be on ``VoD Definition", ``VoD Computation" and the ``User - Zorro - DSP Interface" as that is where the most significant conceptual contributions are. ``Ad Identification \& Categorization" is a necessary component to make Zorro work and so we will describe how to implement it briefly. Our proposed system design will be corroborated with experiments in Section \ref{sec:experiments}.

\subsection{Defining the Value of Data} \label{sec:defining_value_of_data} 
\subsubsection{Problem Setup - CTR Estimation}\label{sec:problem_setup}
We consider the case where there are $m$ users and $n$ advertisers\footnote{A ``user" can also refer to a group of users. Similarly, an ``advertiser" can refer to a category of advertisers.}.
For a given User $i$ and Advertiser $j$, we define the probability of an ad click as $\text{CTR}_{ij}$. We represent these $m \times n$ CTR quantities for each user, advertiser pair through the (unobserved) matrix $\bM \in [0, 1]^{m \times n}$ where $\bM(i, j) = \text{CTR}_{ij}$.

The challenge is that we do not observe the matrix $\bM$. Rather, we observe a sparse, noisy observations of its entries. This is because in reality, most users only see a small percentage of all advertisers; further, for advertisers they do see, they get exposed to their ads only a few times, most likely once. 

We denote our observation matrix as $\bX \in \{[0, 1] \cup \{?\}\}^{m \times n}$. If $\bX_{ij} = ?$, then user $i$ has not been exposed to advertiser $j$. If $\bX_{ij}  \in [0, 1]$, then $\bX_{ij}$ indicates the empirical average of the number of times a user has clicked a particular ad. 

The aim is to infer the CTR for any User $i$ and Advertiser $j$ (i.e., $\bM(i, j)$), given only sparse, noisy observations (i.e., $\bX$).

\subsubsection{Latent Variable Model}\label{sec:LVM}

For any model to effectively learn the underlying CTR given sparse, noisy data, some structure on the matrix $\bM$ must be assumed. A general, flexible way of capturing the underlying structure in social data (e.g. Netflix challenge, product recommendation) is assuming a latent variable model (LVM), arguably the canonical modeling choice from the lens of nonparametric statistics (see cf.~\cite{song2016blind} and references therein). 

Specifically for our case, we assume that the CTR for a given user and advertiser is described by the following latent variable model, $\bM(i, j) = f(\theta_i, \omega_j)$ where $f: \Reals^{d_1} \times  \Reals^{d_2} \to [0, 1]$, $\theta_i \in \Reals^{d_1}$ and $\omega_j \in \Reals^{d_2}$. Here $\theta_i, \omega_j$ refer to multi-dimensional variables that denote latent factors associated with user $i$ and advertiser $j$. $f$ is a latent function that maps $\theta_i$ and $\omega_j$ to a value between $0$ and $1$ denoting the underlying CTR for that user, advertiser pair. 

Note that $f, \theta_i, \omega_j$ are all unobserved (i.e., latent). Instead, recalling Section \ref{sec:problem_setup}, we observe $\bX$ where we model $\bX(i, j)$ as a (independent) random variable such that  $\Ex[\bX(i, j)] = M(i, j)$. From this sparse, noisy $\bX$, we need to estimate $\bM$, which we explain how to do in Section \ref{sec:computing_value_of_data}.

\subsubsection{Definition of Value of Data}\label{sec:definition_value_of_data}

Given the LVM described above, we can now provide a definition of the VoD.

\begin{definition}[Value of Data]\label{def:VoD_definition}
For $i \in [m]$ and $j \in [n]$, the value of User $i$'s data to Advertiser $j$ is given by,
\begin{equation}
	\VoD = |f(\theta_i, \omega_j) - \Ex_\theta[f(\theta, \omega_j)]|
\end{equation}
where $\Ex_\theta[f(\theta, \omega_j)] = \frac{1}{m} \sum^m_{k = 1} f(\theta_k, \omega_j) $
\end{definition} 

In words, this definition states that the value of User $i$'s data to an Advertiser $j$ is the absolute difference between the average CTR for the advertiser across all users (given by $\Ex_\theta[f(\theta, \omega_j)]$) and the CTR for user $i$ (given by $f(\theta_i, \omega_j)$). Though seemingly natural, we consider the definition above as arguably the most important conceptual contribution of this paper and we now make a few remarks regarding it. \\

\vspace{-2mm}

\noindent \textbf{Model Agnostic VoD Definition.} Note that our definition of VoD is not dependent on any advertiser specific model, but rather just on the prediction task itself (in this case CTR estimation), which is what renders it model agnostic. Further, it is worth noting that we can easily replace CTR with any other quantity of interest in a two-sided market (e.g. likelihood of lead conversion in an auto-insurance marketplace, estimated average spend in a retail marketplace). \\

\vspace{-2mm}

\noindent \textbf{Translation of VoD to a Dollar Amount.} To convert this quantity to a dollar amount, we need to multiply $\VoD$ by how much Advertiser $j$ values a marginal increase in accuracy in estimating CTR. For example, it could be be the case that an online healthcare advertiser selling medical drugs values a unit increase in accuracy significantly more than an advertiser selling sports equipment. Hence, the quantity $\VoD$ will need to be scaled to capture this heterogeneity. In Section \ref{sec:dsp_zorro_interface}, we show in detail how such advertiser specific adjustments can be made.  

\subsection{Computing the Value of Data} \label{sec:computing_value_of_data} 

\subsubsection{CTR Estimation via Matrix Estimation}

To make our definition of VoD operational in practice, we need an inference method to estimate CTR without relying on the models of the advertisers, but instead using only observations in terms of historical clicks. Specifically, given $\bX$, estimate $\bM$. This is why we choose matrix estimation, the de facto, non-parametric method to learn the underlying mean of a matrix from sparse, noisy observations. Moreover, it comes with strong theoretical guarantees when a LVM structure is assumed on the underlying mean matrix (cf.~\cite{song2016blind}). Importantly, matrix estimation remains an actively researched area within machine learning with a variety of scalable, production-grade implementations (some standard libraries include \cite{meng2016mllib, abadi2016tensorflow}). 

\vspace{2mm}

\noindent \textbf{Matrix Estimation Overview.} Matrix estimation is the problem of recovering a data matrix from an incomplete and noisy sampling of its entries. Specifically, there exists an underlying $m \times n$ matrix, $\bM$, of interest. We denote the observation matrix as $\bX$, whose entries are noisy versions of $\bM$, i.e., $\bM(i, j) = \Ex[\bX(i, j)]$. In addition to the observations being noisy, it is further assumed that we only observe a small subset of the entries, i.e., each entry $\bX(i, j)$ is observed with probability $\rho \in [0, 1]$. 


This setup has become of great interest due to its connection to recommendation systems, social network analysis, and graph learning (graphon estimation). The key realization of this rich literature is that one can estimate the true underlying matrix from noisy, partial observations by simply taking a low-rank approximation of the observed data. We refer an interested reader to \cite{davenport2016overview} for a broad overview and references therein. 

\begin{definition}[Matrix Estimation]\label{def:matrix_estimation}
A matrix estimation algorithm, denoted as $ME : [0, 1]^{m \times n} \to \Reals^{m \times n}$, takes as 
input a noisy matrix $\bX$ and outputs an estimator $\hM$.
\end{definition}
 
\vspace{-1mm}

\noindent \textbf{CTR estimation via Matrix Estimation.} CTR estimation can be viewed as a special case of matrix estimation. Specifically, $\bX_{ij}$ can be modeled as the empirical average of a Bernoulli random variable with parameter $f(\theta_i, \omega_j)$. Thus we have that $\frac{1}{\rho} \Ex[\bX_{ij}] = f(\theta_i, \omega_j) = \bM_{ij}$ (recall $\rho$ is the fraction of entries observed). This is exactly the setting in which matrix estimation applies. Thus by applying a standard matrix estimation algorithm, such as Singular Value Thresholding (SVT)or Alternating Least Squares (ALS) (cf.~\cite{jain2013low, cai2010singular}), we can reliably infer the underlying $\text{CTR}_{ij}$, simply from sparse, noisy observations (i.e., the small number of click or no click data we have). Recall again that to carry out this procedure, we do not need access to the advertiser's ML model, simply the click data of any user using the Zorro system. 


In Section \ref{sec:ME_experiments} we show on a large real-world online advertising datasets, matrix estimation (using ALS) does reliably infer the underlying CTR from partial, noisy click data (after appropriate aggregation of user and advertiser data). We have out-of-sample performance of $R^2 = 0.58$, which is in line with state-of-the-art performance for matrix estimation. \\

\vspace{-2mm}

\noindent \textbf{Estimated Value of Data Definition.} Analogous to Definition \ref{def:VoD_definition}, we define a notion of VoD that can be empirically measured. We recall certain quantities. Let $\bX$ denote the set of sparse noisy observations of user, advertiser data. Let $\hM =  ME(\bX)$ be the estimated underlying CTR after applying matrix estimation on the sparse, noisy observations $\bX$. We denote $\hf(\theta_i, \omega_j) := \hM_{ij}$. 


\begin{definition}[Estimated Value of Data]\label{def:estimated_VoD_definition}
For $i \in [m]$ and $j \in [n]$, the estimated value of User $i$'s data to Advertiser $j$ is given by,
\begin{equation}
\hVoD = |\hf(\theta_i, \omega_j) - \Ex_\theta[\hf(\theta, \omega_j)]|
\end{equation}
where $\Ex_\theta[\hf(\theta, \omega_j)] = \frac{1}{m} \sum^m_{k = 1} \hf(\theta_k, \omega_j) $
\end{definition} 

In Section \ref{sec:VoD_experiments}, we show on a large real-world online ad click dataset, user's personal data has value (with respect to Definition \ref{def:estimated_VoD_definition}) ranging from 30\% to 69\% depending on advertiser category. Importantly, Zorro allows advertiser's to estimate the VoD for each user and hence purchase data on a query-by-query basis, rather than through inefficient long-term contracts as is done now (details in Section \ref{sec:dsp_zorro_interface}). 

It is also worth highlighting that Definition \ref{def:estimated_VoD_definition} can be operationalized using matrix estimation simply using historical user ad clicks. Thus any DMP (or more generally third party data seller) can perform this function if they collect such data.

\subsection{Ad Identification \& Categorization} \label{sec:ad_collection_categorization}

\subsubsection{Ad Identification}\label{sec:ad_identification}

Recall that the focus of this work is display ads rather than search ads (see Section \ref{sec:introduction}). 

\vspace{2mm}

\noindent \textbf{Definition of Ad.}  Below we describe a straightforward two-level filter system to identify display ads when a user clicks on a link on a website. 

\begin{enumerate}
	\item \textbf{Filter Level 1.} We filter for hyperlinks on a website that redirect to a url external to the website itself. 
	\item \textbf{Filter Level 2.} We parse the redirect url to check for substrings that are almost always present when the external link redirects to ads hosted by third-party servers (e.g. AdChoices, Outbrain). We do this by filtering the redirect url using the industry standard Easylist filter list, which is a set of rules designed by AdBlock (cf.~\cite{AdBlock2017}) to identify ads. 
\end{enumerate}

In Section \ref{sec:ad_experiments}, we show that this simple system effectively identifies whether an external hyperlink is an ad (True Positive and True Negative rate of $83\%$ and $97\%$ respectively). 

\vspace{2mm}

\noindent \textbf{Generalization to Non-Traditional Ads.} It is worth mentioning that companies not traditionally considered advertisers do purchase user data to personalize their website experience for each user (cf.~\cite{EirinakWebMining2003}). For example, an online marketplace company might purchase user browsing data to decide what products to recommend to that user if he or she comes to their homepage. Again users do not get to choose whether this personal data about them is allowed to be sold nor do they derive income from it. We leave it as future work to generalize our definition of ads (with the necessary adjustments made to the entire Zorro system) to just Filter Level 1.

\subsubsection{Ad Categorization}\label{sec:ad_categorization}

\noindent \textbf{Interactive Advertising Bureau (IAB) Overview.} Ad categorization is an important pre-processing step in grouping advertisers together before performing matrix estimation to compute the VoD. Otherwise, the matrix of users and advertisers (i.e., $\bX$) is too sparse and we cannot effectively infer the underlying CTR for a user, advertiser pair. 
We rely on a widely adopted online advertising taxonomy 
\footnote{``IAB Tech Lab Content Taxonomy Version 2.0"} 
published by the IAB (cf.~\cite{IABTaxonomy2018}), which advertisers use to serve more relevant content to users. This context taxonomy has four tiers of classification (e.g. ``Style \& Fashion" $\to$ Women's Fashion $\to$ Women's Clothing $\to$	Women's Business Wear). We focus our categorization system on the top two tiers of IAB categories of where there are $28$ and $341$ respectively. 

\vspace{2mm}

\noindent \textbf{Implementing IAB's Categorization.} Given a website, classifying which category it belongs is a challenging task due to the large number of Tier 1 and 2 categories. To automate the process of classifying websites into IAB categories, we input the text within the body of a webpage into a natural language processing software, called uClassify (cf.~\cite{kaagstrom2013uclassify}), to classify it to an IAB category.

In Section \ref{sec:ad_experiments}, we show this natural language processing system effectively classifies Wikipedia webpages into the correct IAB category, with an average Tier 1 and Tier 2 accuracy of $67\%$ and $53\%$ respectively. Note that a random guess Tier 1 and Tier 2 guess would have an accuracy of approximately $3.6\%$ and $0.3\%$ respectively.

\subsection{User --- Zorro --- Advertiser Interface} \label{sec:dsp_zorro_interface}
In this section, we explain from the advertisers' perspective, how precisely they will interface with Zorro and gain value from it. Recall this part of the system serves two purposes: (1) ensure advertisers pay for user data based only on the increase in accuracy it brings; (2) ensure user data is not inadvertently ``leaked" to advertisers without them paying for it. Our approach consists of two steps: (1) an offline step where advertisers negotiate a contract of how much a marginal increase in accuracy is worth; (2) an online step where advertisers query Zorro for user data, and pay for it based on the estimated increase in accuracy they will receive from purchasing it. 

\vspace{2mm}

\noindent \textbf{Offline Contract.} Recall from Section \ref{sec:definition_value_of_data} that each advertiser can have a very different value for an increase in accuracy (consider an online advertiser selling medical drugs vs. sports equipment - how much each values an unit increase CTR can by vastly varying). Thus the estimated VoD (given by Definition \ref{def:estimated_VoD_definition}) needs to be scaled accordingly. Doing this scaling is precisely the purpose of the offline contract between a DSP and Zorro. We describe the contract below.

Recall from Section \ref{sec:ad_categorization} that before doing matrix estimation, we group advertisers together by IAB category (e.g. ``Disease and Conditions", ``Sports Equipment"). Since we estimate the VoD per advertising category, it follows that this offline contract must be defined per IAB category as well. Let there be $C$ such categories. Then for advertiser $j \in [M]$ and category $c \in [C]$, we define a function $p_j^c : [0, 1] \to \Reals^+$. This function, $p_j^c$ summarizes how much advertiser $j$ values an increase in accuracy for estimating CTR for category $c$. 
 
For example, a sports equipment advertiser might value a $10\%$ increase in accuracy in the category ``Sports Equipment" to be worth $\$0.1$ while may value a $20\%$ increase in the category ``Disease and Conditions" to only be worth $\$0.001$. The functions $p_j^c$ capture this heterogeneity. Thus a negotiated contract between Zorro and an advertiser $j$ will be the set $P_j := \{p_j^c: \ \text{for} \ c \in [C]\}$

\vspace{2mm}

\noindent \textbf{Real-Time Interface.} In the example below, let there be $K$ advertisers denoted $\{\text{AD}_1, \dots, \text{AD}_K\}$ that have negotiated an offline contract with Zorro. We describe the proposed sequence of interactions that will occur between advertisers and Zorro, each time a user on Zorro loads a publisher's webpage. 

Whenever a user sends a webpage request, it is routed through a Zorro proxy server and is anonymized by replacing the User ID (i.e. user cookies) by a random ID (i.e. random cookie). Advertiser's that have signed a contract with Zorro can query the system to ask for user data before deciding what bid to make. They then pay for this data based on the estimated VoD and the offline contract. A key point to highlight is that a new random ID is generated each time a user loads a webpage, which is what anonymizes the user request.  

Below is the pseudocode for this procedure. Refer to Figure \ref{fig:RTB_overview} for a step-by-step overview of how RTB would work with and without Zorro (adapted from \cite{zhang2014real}).

\begin{figure*}
\centering

\subfigure[]{%
\label{fig:RTB_overview_1}%
\includegraphics[width=0.48\linewidth,height=3.2cm]{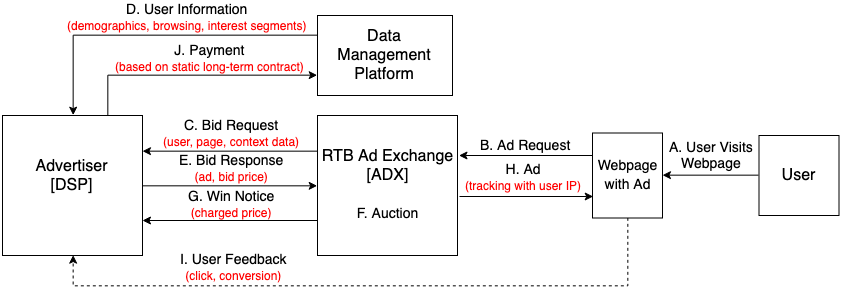}}%
\vspace{-3mm}
\hfill
\subfigure[]{%
\label{fig:RTB_overview_2}%
\includegraphics[width=0.48\linewidth,height=3.2cm]{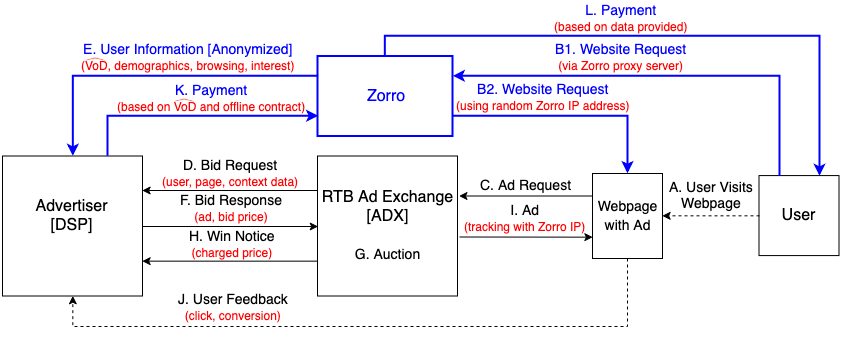}}%
\vspace{-1mm}
\caption{Figure \ref{fig:RTB_overview_1} - how RTB works with current DMPs. Figure \ref{fig:RTB_overview_2} - how we propose RTB work with Zorro.}
\label{fig:RTB_overview}
\end{figure*}

\vspace{3mm}

\hrule  
\begin{enumerate} 
	\item A random Zorro ID cookie is generated and linked with User $i$'s actual ID cookie. 
	\item User $i$'s with Zorro extension has their webpage request routed through a Zorro proxy server. 
	\item User $i$'s loads a webpage.
	\item The publisher pings a RTB exchange for ad requests and the RTB forwards the requests to advertisers.
	\item $\{\text{AD}_1, \dots, \text{AD}_K\}$ query Zorro to check if User $i$'s ID cookie belongs to Zorro \footnote{The random Zorro ID cookie could contain a Zorro specific numeric signature to signify that user has the Zorro extension installed}
	\item If User $i$'s ID cookie belongs to Zorro, then following two actions occur
	\begin{enumerate}
		\item Zorro applies matrix estimation and sends $\{\text{AD}_1, \dots, \text{AD}_K\}$ following two signals: (i) for each $\text{AD}_k$, the estimated VoD of User $i$'s data; (ii) User $i$'s actual personal data (e.g. demographic, interest segments, browsing history).
		\item $\{\text{AD}_1, \dots, \text{AD}_K\}$ pays Zorro based on offline contract defined above. Specifically $\text{AD}_j$ pays,
		 \[\max_{c \in [C]}\bigg(p_j^c\Big(\hVoD\Big)\bigg).\]
	\end{enumerate}
	\item A new random Zorro ID cookie is generated and linked with User $i$'s actual ID cookie.
\end{enumerate}
\hrule 

\vspace{4mm}

\noindent \textbf{Why User Data is Not ``Leaked"?} By refreshing a new random Zorro ID cookie every time a user loads a webpage, we ensure that to an advertiser who has not negotiated a contract with Zorro, it seems like a random user is loading the webpage. A crucial benefit of this architecture is that user data is never permanently ``leaked" as the random Zorro ID cookie is constantly refreshed. Thus if a user who has installed Zorro decides to go incognito at some point and no longer receive money for providing their personal data, they easily can.

\subsubsection{Model Agnostic Data Contracts}

The problem of companies not wanting to share their models or their data before a transaction has happened clearly extends far beyond online advertising. This is why so many firms are stuck in inefficient data contracts, where they purchase data based on a long-term agreement or a coarse metric such as the number of API calls they make to a data seller. 

That is why we believe our proposed approach to designing data contracts can serve as a general-purpose, empirically auditable system for companies to reason about the VoD when sourcing from external vendors. In essence, Zorro can serve as a base to allow data buyers to pay based only on the estimated \textit{increase in accuracy for a clearly defined prediction task}, when purchasing a new dataset from third-parties.



\section{Experiments}\label{sec:experiments}

The experiments are organized to answer four set of questions: 
\begin{enumerate}
	\item \textbf{Matrix Estimation.} How accurately matrix estimation is able to estimate the underlying CTR?
	\item \textbf{VoD.} How much variation in VoD for user data is there depending on an advertiser's category?
	\item \textbf{Ad Data Collection.} How accurately can we identify and categorize ads?
	\item \textbf{User Data Protection.} Can we reliably prevent user data from being ``leaked" to advertisers?
\end{enumerate}

\subsection{Matrix Estimation Experiments}\label{sec:ME_experiments}

\noindent \textbf{Matrix Estimation Algorithms Applied.} We focus on two well-studied matrix estimation algorithms, Alternating Least Squares (ALS) and Singular Value Thresholding (SVT). See \cite{jain2013low, cai2010singular} for a detailed description of these algorithms and theoretical guarantees of convergence and quality of the final estimate. For ALS, we rely on a standard PySpark implementation (cf.~\cite{meng2016mllib}) and for SVT, we use the NumPy package (cf.~\cite{van2011numpy}) to perform the singular value decomposition. Lastly, we implement a hybrid two-step algorithm that first does SVT on the observation matrix and uses the outputted matrix as a ``warm start" for ALS. 

\vspace{2mm}

\noindent \textbf{Metric Used to Measure Accuracy.} We measure quality of performance using $R^2$
\footnote{For a set of values $\{y_1, \dots, y_n\}$ and estimates $\{\hat{y}_1, \dots, \hat{y}_n\}$, $R^2= 1 - \frac{\text{SS}_{\text{Var}}}{\text{SS}_{\text{Res}}}$, where $\text{SS}_{\text{Var}} = \sum_{i =1}^n (y_i - \bar{y})^2$,  $\text{SS}_{\text{Res}} = \sum_{i =1}^n (\hat{y}_i - \bar{y})^2$, and $\bar{y} = \frac{1}{n} y_i$}. 

%

\subsubsection{Avito Dataset}\label{sec:avito_dataset}

\noindent \textbf{Description of Dataset Used.} We test our system on the Avito Context Ad Clicks dataset. To the best of our knowledge, it is the largest and most comprehensive publicly available ad click dataset. It can be found on Kaggle (cf.~\cite{KaggleAvito2015}).  Avito is the largest online marketplace in Russia with 70 million unique monthly visitors. Another reason we use this dataset is that its focus is contextual ads (i.e. banner ads), which is also the focus of our work.

There are $190,157,735$ context ad views in the dataset, out of which $1,146,289$ led to clicks. For each view, the dataset includes metadata about the user (e.g. geographic location, parameters of the search) and the ad itself (e.g. category of the ad, historical CTR), along with whether the view led to a click or not.

\vspace{2mm}

\noindent \textbf{Grouping Users \& Advertisers.} We group users by their geographical location (there are 3431 unique location IDs) and ads by their category (there are 31 unique category IDs). We choose location as it is a standard feature used in CTR estimation, and related tasks. However, many other user features can and should be incorporated in practice. Indeed, our experiments are meant to showcase accurate CTR estimation is possible by simple groupings of users. Similarly, in practice advertisers are commonly grouped by category (see Section \ref{sec:ad_categorization}) and hence we do so as well. 

\vspace{2mm}

\begin{figure*}[]%
\centering

\subfigure[]{%
\label{fig:ME_data_quality_1}%
\includegraphics[width=0.45\linewidth,height=4.5cm]{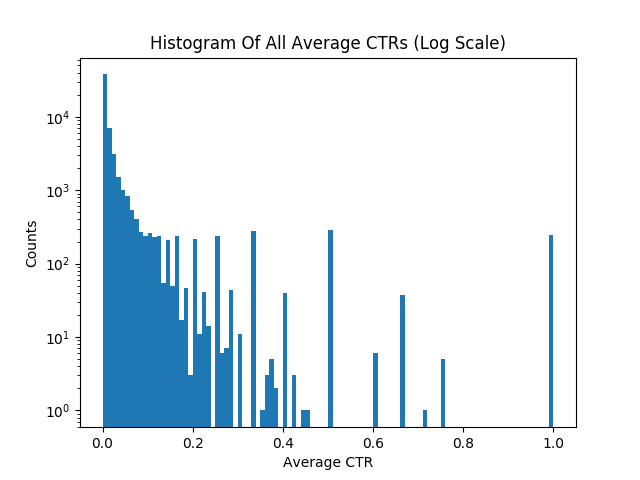}}%
\hfill
\subfigure[]{%
\label{fig:ME_data_quality_2}%
\includegraphics[width=0.45\linewidth,height=4.5cm]{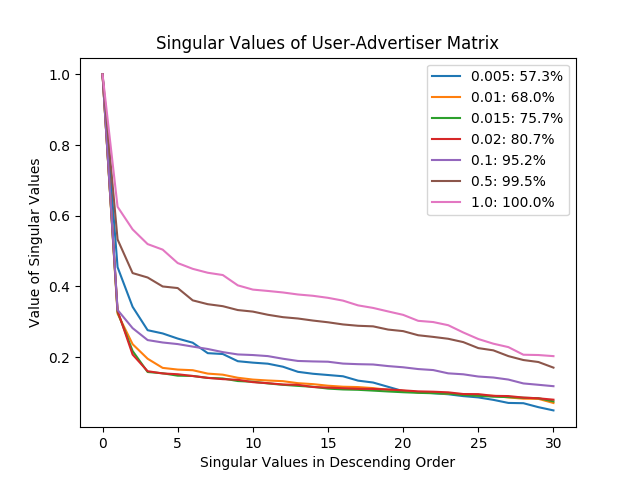}}%
\vspace{-3mm}
\caption{In Figure \ref{fig:ME_data_quality_1}, we plot the histogram of CTR for different users. In Figure \ref{fig:ME_data_quality_2}, we plot singular values at different CTR thresholds and percentage of data retained.}


\subfigure[]{%
\label{fig:ME_percent_matrix_filled_1}%
\includegraphics[width=0.33\linewidth,height=4cm]{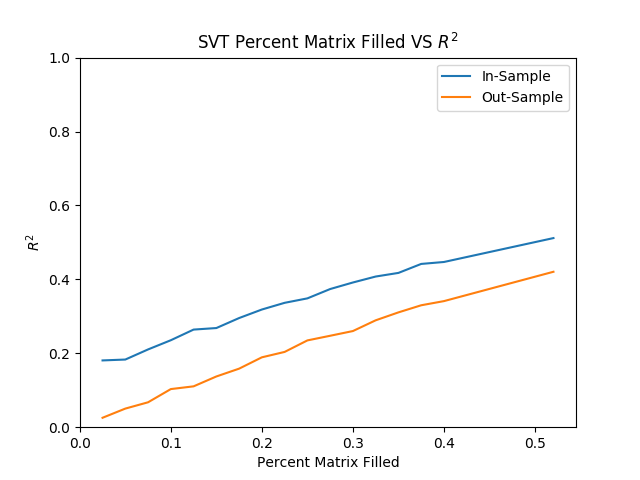}}%
\hfill
\subfigure[]{%
\label{fig:ME_percent_matrix_filled_2}%
\includegraphics[width=0.33\linewidth,height=4cm]{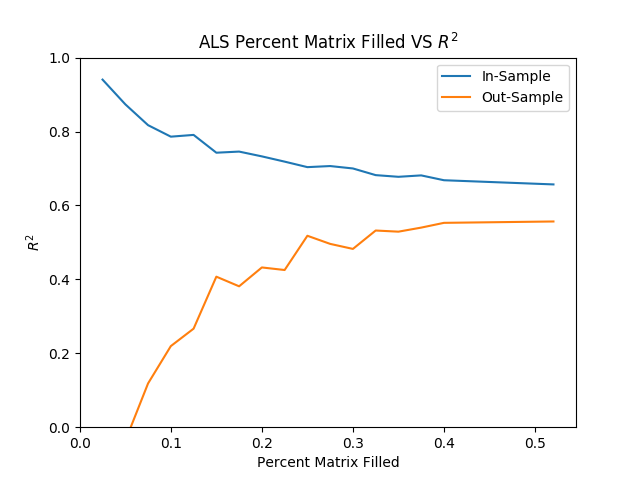}}%
\hfill
\subfigure[]{%
\label{fig:ME_percent_matrix_filled_3}%
\includegraphics[width=0.33\linewidth,height=4cm]{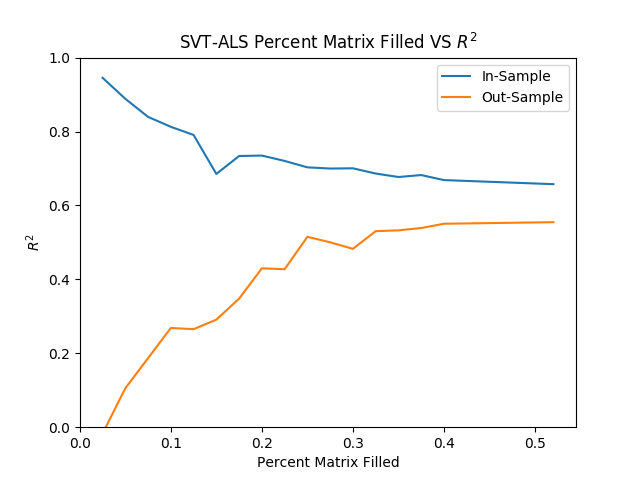}}%
\vspace{-3mm}
\caption{$R^2$ vs. \% Matrix Filled for SVT, ALS and SVT +  ALS. In-Sample and Out-of-Sample performance plotted.}


\subfigure[]{%
\label{fig:ME_data_points_1}%
\includegraphics[width=0.33\linewidth,height=4cm]{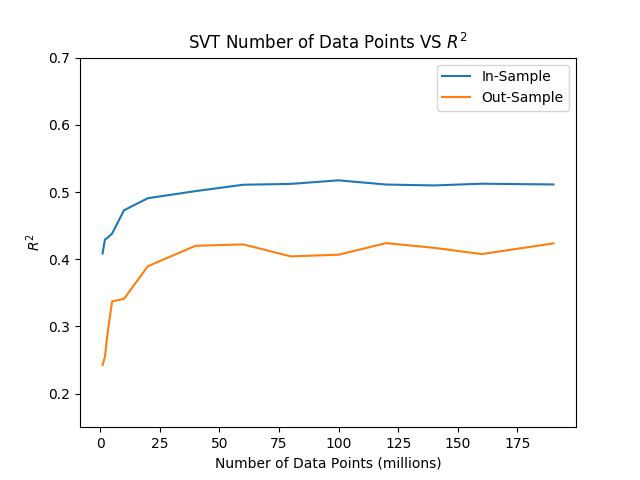}}%
\hfill
\subfigure[]{%
\label{fig:ME_data_points_2}%
\includegraphics[width=0.33\linewidth,height=4cm]{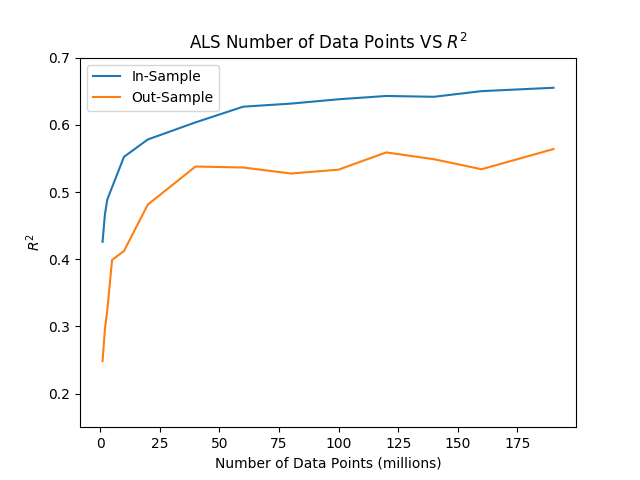}}%
\hfill
\subfigure[]{%
\label{fig:ME_data_points_3}%
\includegraphics[width=0.33\linewidth,height=4cm]{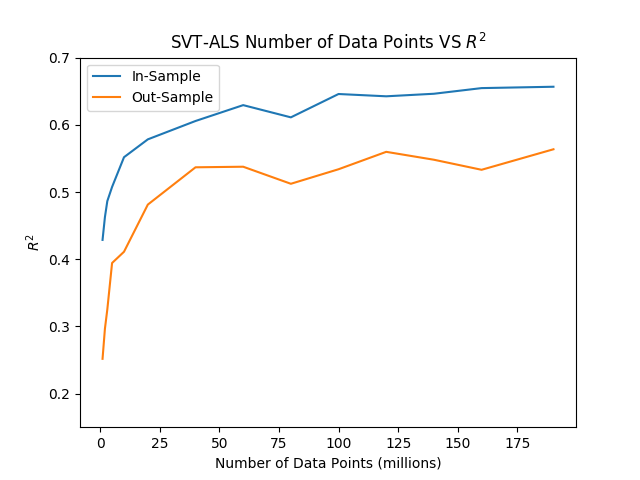}}%
\vspace{-3mm}
\caption{$R^2$ vs. \# Data Points used for SVT, ALS and SVT +  ALS. In-Sample and Out-of-Sample performance plotted.}



%
\subfigure[]{%
\label{fig:ME_signal_vs_noise_1}%
\includegraphics[width=0.33\linewidth,height=4cm]{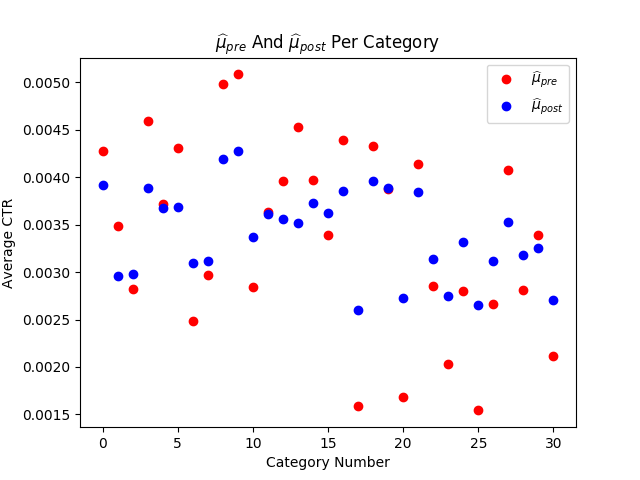}}%
\hfill
\subfigure[]{%
\label{fig:ME_signal_vs_noise_2}%
\includegraphics[width=0.33\linewidth,height=4cm]{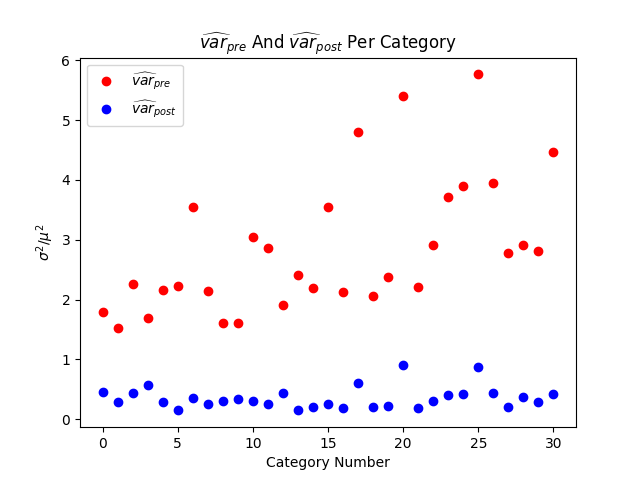}}%
\hfill
\subfigure[]{%
\label{fig:ME_signal_vs_noise_3}%
\includegraphics[width=0.33\linewidth,height=4cm]{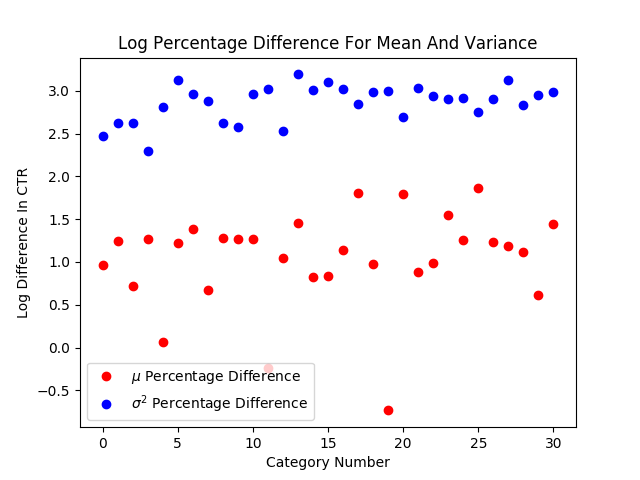}}%
\vspace{-3mm}
\caption{Figures \ref{fig:ME_signal_vs_noise_1}-\ref{fig:ME_signal_vs_noise_2} - Plot of difference in empirical mean and variance per category pre- and post-ME respectively. Figure \ref{fig:ME_signal_vs_noise_3} - \% Difference in $\hat{\mu}$ and $\widehat{\text{var}}$ per category plotted on log scale.}
\end{figure*}

\noindent \textbf{Data Preprocessing.} In Figure \ref{fig:ME_data_quality_1}, we plot a histogram of empirical CTR for each user in the Avito dataset (on a log scale). We see a large number of user CTRs that are unreasonably high (e.g. there is a spike at $1.0$). Indeed, a known issue in such datasets is ``click-bots" which are programs that automatically click on ads to increase revenue in a fraudulent manner. That is why we filter out all users beyond a certain threshold, specifically $0.02$. We do so because according to historical Google AdWords data, ``human" CTR for different advertiser categories ranges between $0.0035$ and $0.02$. (cf.~\cite{Wordstream2016}). 


This choice of threshold is justified empirically through Figure \ref{fig:ME_data_quality_2}. We create a matrix of users and advertisers (grouped according to location and category respectively), take its singular value decomposition, and then plot the singular values of this matrix for different thresholds. If we don't threshold (i.e. keep the threshold at $1.0$), then the spectrum decays very gradually. However, as we decrease the threshold, the spectrum decays progressively quicker, and the low-rank structure we expect to see emerges (see Section \ref{sec:computing_value_of_data}). The optimal threshold is approximately at $0.02$, and at this threshold, we still retain $81\%$ of the original data.

Figure \ref{fig:ME_data_quality_2} is encouraging as it gives credence to the theoretical model laid out in Section \ref{sec:LVM}. A LVM suggests that the underlying matrix, $\bM$ ought to be approximately low-rank. And indeed, Figure \ref{fig:ME_data_quality_2} shows that the underlying data is approximately rank 2! This justifies the LVM and provides a structural reason for the success of the matrix estimation. 

\subsubsection{Matrix Estimation Results}\label{sec:ME_experiments_results}

\noindent \textbf{Hyper-parameter Selection.} For both ALS and SVT, there is one major hyper-parameter to choose, the number of singular values to keep. From Figure \ref{fig:ME_data_quality_2}, we see that the top two singular values contain most of the ``signal" and hence we choose to keep two singular values. This is further justified through cross-validation. In PySpark's implementation of ALS, a regularization parameter must also be chosen, which we set to be $0.01$, again using cross-validation.

\vspace{2mm}

\noindent \textbf{CTR Estimation via Matrix Estimation Experiments.} Figures \ref{fig:ME_percent_matrix_filled_1}-\ref{fig:ME_data_points_3} show how the quality of matrix estimation changes as we vary: (1) the percentage of entries observed; (2) the number of data points used. 

For Figures \ref{fig:ME_percent_matrix_filled_1}-\ref{fig:ME_data_points_3}, we keep the percentage of data withheld for out-of-sample testing to be fixed at $20\%$. We see that $R^2$ performance of the various algorithms improves steadily as the fraction of data observed and the number of data points increases (with ALS outperforming SVT). However, this improvement levels off after approximately $40\%$ of the data is observed and $50$ million data points are used. This indicates two takeaways with regards to accurate CTR estimation: (1) some amount of grouping of users and advertisers is an important pre-processing to ensure the resulting matrix is not too sparse; (2) our system needs to collect approximately on the order of tens of millions of user data to apply matrix estimation effectively.

%
\vspace{2mm}

\noindent \textbf{Signal vs. Noise Thresholding via Matrix Estimation.} From Section \ref{sec:computing_value_of_data}, we know that matrix estimation should help uncover the underlying mean of the CTR matrix, $\bM$, while reducing the noise as long as the underlying matrix is low-rank (which we verified through Figure \ref{fig:ME_data_quality_2}). In particular for the Avito dataset, we expect that the average CTR per advertiser category should remain approximately the same both before and after applying matrix estimation
\footnote{There are a large number of user location id's (3434 of them). Thus the pre-$\text{ME}$ estimate of CTR per advertiser category should be relatively stable.}. 
However, the variance in CTR per advertiser category should drop significantly. 

We recall the notation used in Section \ref{sec:computing_value_of_data}. Let 
\[
\hmupre^j := \frac{1}{\text{\# observed}}\sum_{i=1}^m \bX_{ij} \mathbbm{1}(\bX_{ij} \neq ?), \ \ \ \hmupost^j := \frac{1}{m}\sum_{i=1}^m \hM_{ij}.
\]
In words, $\hmupre^j, \hmupost^j$ are the pre-$\text{ME}$ and post-$\text{ME}$ empirical average CTRs for each advertiser category $j$ in the Avito dataset. Let,
\begin{align*}
\hvarpre^j &:= \frac{1}{\text{\# observed}} \sum_{i=1}^m  (\bX_{ij} \mathbbm{1}(\bX_{ij} \neq ?) - \hmupre^j)^2, \\
\hvarpost^j &:= \frac{1}{m}\sum_{i=1}^m (\hM_{ij} - \hmupost^j)^2.
\end{align*}
In words, $\hvarpre^j, \hvarpost^j$ are the pre-$\text{ME}$ and post-$\text{ME}$ empirical variance in CTR for each advertiser category $j$ in the Avito dataset.

In Figures \ref{fig:ME_signal_vs_noise_1}-\ref{fig:ME_signal_vs_noise_3}, we plot the normalized differences in mean and variance per advertiser category
\footnote{We normalize by $\hmupre^j$ to ensure $\hmudiff$ can be appropriately compared between categories. For example, a $0.1$ difference is significant if the underlying average CTR is $0.01$, but not as much if it is $0.5$. Same reasoning applies for $\hvardiff$.},
\[
\hmudiff := \frac{1}{\hmupre^j}|\hmupre^j -  \hmupost^j|, \ \ \ \hvardiff := \frac{1}{(\hmupre^j)^2}|\hvarpre^j - \hvarpost^j|.
\]
We see from Figure \ref{fig:ME_signal_vs_noise_3} that $\hmudiff$ is of the order $0.2\%-73\%$, while $\hvardiff$ is of the order $200\%- 1550\%$. Thus as desired, $\hmudiff$ is on average approximately three orders of magnitude smaller than $\hvardiff$, indicating that matrix estimation does indeed effectively retain signal and threshold out noise.

\vspace{2mm}

\noindent \textbf{Summary.} In summary, ALS outperforms SVT (the two-step procedure of applying SVT and then ALS gives comparable performance). The best $R^2$ achieved is 0.58; This is in-line with or better than state-of-the-art $R^2$ performance for well-studied datasets such as Movielens, which have performance in the range of $0.3-0.4$ (cf.~\cite{karandikarcse}). Further, our experiments show that matrix estimation does effectively retain signal and threshold out noise.

Thus matrix estimation (specifically, ALS) can be used as an accurate, model-agnostic method to do CTR estimation as long as Zorro has access to sufficient amount of user data, and careful grouping of users and advertiser is done.

\subsection{Value of Data Experiments}\label{sec:VoD_experiments}

An important question to ask is if with respect to Definition \ref{def:estimated_VoD_definition} of VoD, whether an advertiser finds value in a user's personal data? 
Intuitively, if the CTR distribution for a particular advertiser category is tightly concentrated around its mean, then acquiring user data is not as important for that advertiser (as there is not much variation in CTR across users). In contrast, if the CTR distribution for an advertiser category is diffuse, then advertisers are incentivized to acquire personal user data as it helps them identify which segment of the distribution a user lies in. 

\vspace{2mm}

\begin{figure}[h!]%
\centering
\subfigure[]{%
\label{fig:VoD_variation_1}%
\includegraphics[width=0.9\linewidth,height=4.5cm]{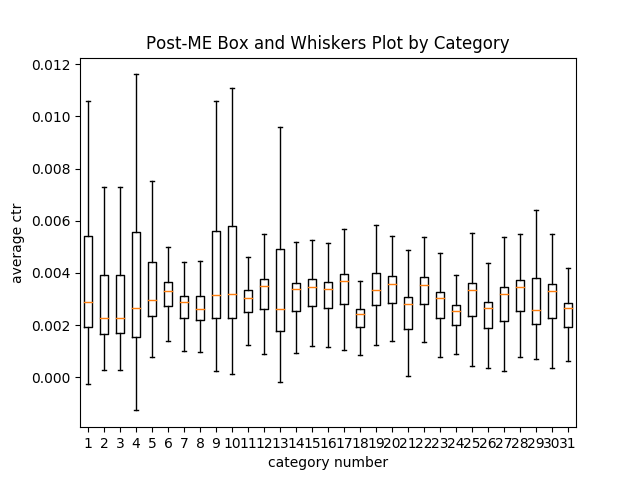}}%
\vfill
\subfigure[]{%
\label{fig:VoD_variation_2}%
\includegraphics[width=0.9\linewidth,height=4.5cm]{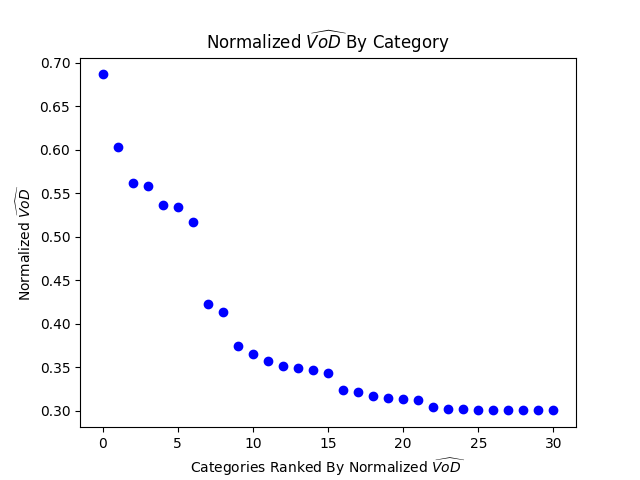}}%
\vspace{-3mm}
\caption{Figure \ref{fig:VoD_variation_1} - Box and Whisker's plot of CTR distribution per category. Figure \ref{fig:VoD_variation_2} - Plot of normalized average $\widehat{\text{VoD}}$ per category.}
\end{figure}

\noindent \textbf{Large Variation in VoD Across Advertiser Categories.} We plot the distribution of CTR per category using a Box-and-Whisker's plot in Figure \ref{fig:VoD_variation_1}, and see instances of both cases above. For example, advertisers in Category 18 have relatively lower VoD as the CTR distribution is tightly concentrated, while advertisers in Category 4 have quite a large VoD as the CTR distribution is quite diffuse. Thus we can visually tell that advertiser's do indeed have a different VoD depending on their category, giving credence to Definition \ref{def:estimated_VoD_definition}. 

We can summarize these visual findings through the following quantity,
\[
\hVoDUser_j := \frac{1}{\hmupre^j}\bigg(\frac{1}{m}\sum_{i =1}^m \hVoD \bigg).
\]
In words, $\hVoDUser_j$ is the average VoD Advertiser $j$ experiences per user
\footnote{Again, like in Section \ref{sec:ME_experiments_results}, we normalize by $\hmupre^j$ to ensure we can appropriately compare between categories.}. 
In Figure \ref{fig:VoD_variation_2}, we plot $\hVoDUser_j$ for the various Avito categories. We see that $\hVoDUser_j$ ranges from 30\% to 69\% depending on advertiser category, indicating that there is large variation in VoD depending on which category an advertiser belongs to.

\vspace{2mm}

\noindent \textbf{Query-by-Query Estimation of VoD.} It is important to highlight that in our proposed system, Advertiser $j$ can compute the VoD on a query-by-query basis. In other words, each time User $i$ loads a webpage, Advertiser $j$ can compute $\hVoD$. This is contrast with how advertiser's and third-party data providers engage currently, where advertisers pay based on long-term contracts or a coarse metric such as number of API calls made.

\vspace{2mm}

\noindent \textbf{Estimating Dollar Value of VoD.} Ideally we would like to translate the estimated VoD to a dollar amount per advertiser category. Unfortunately, the Avito advertiser categories are anonymized and so instead, we give a conservative estimate. From Section \ref{sec:introduction}, the size of the (rapidly growing) display advertising market is USD 54 Billion. Since the minimum $\hVoDUser_j$ is estimated to be $30\%$ (in line with previous work, see Section \ref{sec:related_work}), it follows that if Zorro prevents advertisers from getting user data without paying for it, the minimum loss of value for advertisers is on the order of USD 16 Billion.

\subsection{Online Ad Experiments}\label{sec:ad_experiments}

In this section, we test how well we identify and categorize online ads by tying together open source software packages.


%
%

\subsubsection{Online Ad Identification}

\noindent \textbf{Overview of Ad Identification Module.} Recall from Section \ref{sec:ad_identification} that the main filter we use on the redirect url is EasyList, the same filter set used by AdBlock. In addition, we manually add keywords such as “doubleclick” and “criteo” that commonly occur when a user has clicked on an ad.

\vspace{2mm}

\noindent \textbf{Overview of Ad Identification Experiment.} To evaluate this simple system, we conducted a crowdsourced experiment. The experiment consisted of iterating through Alexa’s top 500 global sites (cf.~\cite{AlexaTop5002018}) and clicking ten unique ads on the first 30 valid sites (websites that were not in English or had no ads were skipped). If there were not ten unique ads on the websites landing page, then ads on different webpages under the some domain were searched for. Additionally, hyperlinks that were not ads were also clicked to test whether our system could also identify non-ads accurately.

\vspace{2mm}

\noindent \textbf{Result of Ad Identification Experiment.} The result of the above crowdsourced experiment are shown below. 

\begin{table}[h!]
  \begin{center}
    \begin{tabular}{l|c|r} 
       & \textbf{True Ad} & \textbf{Not Ad}\\
      \hline
      \textbf{Classified Ad} & 239 (83\%) & 49 (17\%)\\
      \textbf{Classified Not Ad} & 2 (97\%) & 68 (3\%)\\
   \end{tabular}
  \end{center}
   \label{tab:AD_identification_table}
\end{table}

We have a True Positive rate of $83\%$ and a True Negative Rate of $97\%$, indicating that our system can reliably identify whether a hyperlink on a webpage is an ad or not. Additionally, a large portion of the ads we could not identify were because they did not go through an ad exchange (i.e. have a redirect url that spawned a tab in a browser). If we do not include such ads, then our True Positive rates goes up to $88\%$. Such ads are the result of direct negotiations between a website and advertisers. Thus they are not of relevance to our system as they do not go through the RTB ecosystem, and third-party data is not purchased by advertisers in such instances.

\subsubsection{Online Ad Categorization}

\noindent \textbf{Overview of Ad Categorization Module.} Recall from Section \ref{sec:ad_categorization} (and verified in Section \ref{sec:ME_experiments_results}) that ad categorization is an important pre-processing step to group advertiser's together to get good out-of-sample performance in CTR estimation. As detailed in Section \ref{sec:ad_categorization}, we use the industry standard IAB taxonomy to categorize advertisers. To do so, we use the open-source natural language processing software uClassify to predict a webpage's IAB category based on the text in the webpage's body. We focus on the top two Tiers of the IAB taxonomy as the categorization is sufficiently detailed. It is straightforward to extend to the bottom two tiers if required. 

\vspace{2mm}

\noindent \textbf{Overview of Ad Categorization Experiment.} Unfortunately, there do not exist good online datasets with IAB categorizations of websites that we can use to test the accuracy of our system. Thus we designed a proxy experiment using Wikipedia webpages as follows. Wikipedia provides a detailed hierarchal categorization of its own webpages. Thus for each IAB Tier 1 category ($28$ in total), we randomly sample two Tier 2 categories within it and then define a mapping from these IAB Tier 2 category to a near identical Wikipedia category. For example, the IAB category "Travel Type" was mapped to the Wikipedia category "Types of Travel". If a close Wikipedia category was not found, we resample an IAB Tier 2 category such that a matching Wikipedia is found.

For each Wikipedia category selected, we chose up to 100 articles under that category. If there were not 100 articles for a given category (for many there were not), we iterated through the subcategories of that category, and randomly sampled articles from subcategories until we reached a 100. We then ran our classifier on the text from those articles. 

\vspace{2mm}

\begin{figure}[ht]
\centering
\subfigure[]{%
\label{fig:AD_cat_1}%
\includegraphics[width=0.9\linewidth,height=4.5cm]{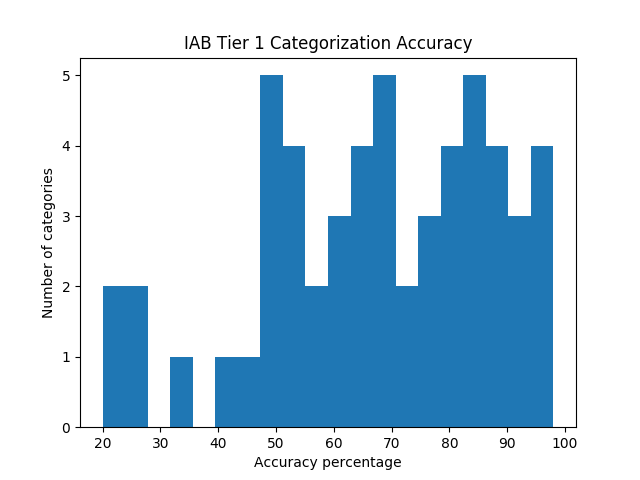}}%
\hfill
\subfigure[]{%
\label{fig:AD_cat_2}%
\includegraphics[width=0.9\linewidth,height=4.5cm]{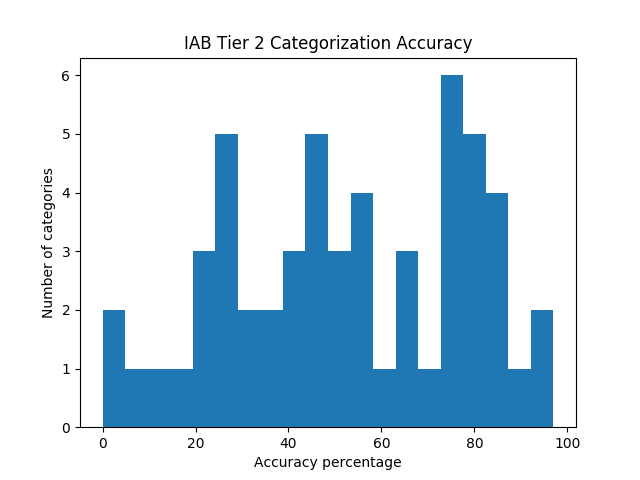}}%
\caption{Average uClassify classifier accuracy on IAB Tier 1 and Tier 2 categories is 67.36\% and 52.75\% respectively.}
\end{figure}

\noindent \textbf{Result of Ad Categorization Experiment.} The results of the above  experiment are shown in Figures \ref{fig:AD_cat_1} and \ref{fig:AD_cat_2}. For Tier 1 categories, the classification accuracy was 67.36$\%$ and for Tier 2 categories, the classification accuracy was 52.75$\%$. This is consistent as correctly classifying a Tier 1 category is a necessary requirement to correctly classify a Tier 2 category.

It is worth noting that these results are far better than if we random selected a category. Recall that there are 28 IAB Tier 1categories and 341 Tier 2 categories and so a random classification would give $3.6\%$ and $0.3\%$ accuracy respectively. 

Furthermore, it is likely that our system performs better than our tests indicate for the following two reasons. (1) It was often that case that articles from subcategories of the selected Wikipedia category were somewhat unrelated the further down in the Wikipedia hierarchy the subcategory was. For example, the category "Environment" was mapped to a Wikipedia subcategory "Underground Laboratories". In such mappings, correct classification is not possible. (2) If an outputted IAB category was incorrect, it was still  often strongly related to the correct one. For example, articles on children’s films would be classified as family or fantasy films. 

Considering these relatively benign sources of error, and the large number of Tier 1 and Tier 2 categories, our system reliably categorizes an advertiser to a relevant IAB category, based on the advertiser's webpage.

\vspace{-2mm}

\subsection{User Data Protection Experiments}\label{sec:protecting_user_data}

Recall from Section \ref{sec:dsp_zorro_interface} that for our system, it is essential that personally identifiable information (PII) of a user on the Zorro system, is not inadvertently ``leaked" to advertisers. We begin by giving a brief overview of how advertisers track users across the web and then describe an experiment to verify that user data is indeed not leaked to an advertiser.

\vspace{2mm}

\noindent \textbf{Tracking Techniques.} As users surf the web, advertisers track users through three main methods (cf.~\cite{mayer2012third}). 

(1) Third Party Cookies: Primary method of tracking. Cookies on a user's browser not placed there by the primary website he/she is visiting
\footnote{First-party cookies are placed on a user's browser by the primary website. Used to store site specific settings such as login information.}.
Meant solely for tracking users across web. Hence we identify and remove such cookies by checking that they are unrelated to the primary website. 

(2) Cache: Images and other website-specific information are stored on the browser as cookies. Meant for more efficient page loads as website information need not be retrieved from the server. However, websites can encode information in such cookies to identify and track users. Easily easily solved by clearing the cache periodically. However from the experimental results, this is not currently necessary as websites do not tend to forward this information on to advertisers.

(3) Flash Cookies: Cookies stored in flash player plugins (e.g. Adobe Flash Player). Flash cookies are harder to clear without leading to a significantly poorer user experience; specifically, clearing flash cookies would require clearing data from every plugin a user has installed. Again as with (2), from experimental results, this is not currently necessary as a minimal number of advertisers go through such a mechanism.

\vspace{2mm}

\noindent \textbf{Overview of User Data Protection Experiment.} We design a simple test to verify whether our system prevents inadvertent ``leakage" of user data to advertisers. We do so by clicking on an ad on an initial website, and counting the number of times that ad is repeated on subsequent websites. We do this experiment with: (1) clearing no cookies; (2) blocking only third-party cookie blocking; (3) blocking all types of cookies listed above. We automate this experiment by using software packages Selenium and Pyautogui to automatically click on ads on the top 500 websites listed by Alexa (cf. \cite{AlexaTop5002018}). On each website, our software clicks every ``iframe" element and uses our ad identification module to classify if is is an ad or not. 


\vspace{2mm}

\noindent \textbf{Result of User Data Protection Experiment.} See Table \ref{tab:USER_hashing} for results of the above experiment. As desired, we see that when we simply block third-party cookies, we reduce the percentage of repeat ads show from $29\%$ to $9\%$ as compared to if there was no blocking. By blocking all cookies listed above, it does not affect the number of repeat ads seen, indicating that blocking Cache and Flash cookies is currently less important.

\begin{table}[h!]
  \begin{center}
    \begin{tabular}{l|c|c|c} 
       & \textbf{Zorro Off} & \textbf{3rd Party} & \textbf{All} \\
      \hline
      \textbf{Total Ads Clicks} & 103 & 107 & 107 \\
      \textbf{\# Repeat Ads} & 30 (29\%) & 10 (9\%) & 10 (9\%)\\
   \end{tabular}
  \end{center}
   \label{tab:USER_hashing}
\end{table}

\vspace{-10mm}

\section{Forward Looking Intent}\label{sec:forward_intent}

\textbf{Explicit User Intent in Display Advertising.} A large loss of value in the current ecosystem is that user intent is not captured. For example, it is common in online advertising that once a user has made a retail purchase, ads for the purchased item are repeatedly shown to the user after the fact. This is clearly a source of large loss of revenue for advertisers and a great irritant to users. 

In a future iteration of Zorro, we intend to provide an API to users that let them indicate forward-looking intent for the category of ads they would like to see. Below we extend the LVM described in Section \ref{sec:LVM} to incorporate user intent. Further, we describe an algorithm to estimate CTR that exploits the additional structure when users provide intent. We then verify that our proposed algorithm does indeed exploit the additional structure when estimating CTR

\vspace{2mm}

\noindent \textbf{CTR Estimation with Intent.} Let the set of categories a user can provide intent for be indexed by $[k] := \{1, \dots, k\}$. Then we can model user intent by generalizing $\bM$ to an order-3 tensor, $\bT \in [0, 1]^{m \times n \times k}$. Here the $l$-th, $m \times n$ ``slice"  of $T$ is a matrix of users and advertisers (appropriately grouped) \textit{conditioned} on user intent being $l \in [k]$. As before, we get noisy, sparse observations from $\bT$, which we denote as $\bW \in  \{[0, 1] \cup \{?\}\}^{m \times n \times k}$. Analogous to the matrix case, if $\bW(i, j, l) = ?$, then User $i$ has not been exposed to Advertiser $j$, while providing intent $l$. And if $\bW(i, j, l)  \in [0, 1]$, then $\bW_{ij}$ indicates the empirical average of the number of times a user has clicked a particular ad while providing a specific intent. The aim is estimating $\bT$ from sparse, noisy observations in $\bW$.

\vspace{2mm}

\noindent \textbf{LVM for CTR estimation with Intent.} Recall from Section \ref{sec:LVM} that we impose a LVM on the underlying mean matrix $\bM \in [0, 1]^{m \times n}$, where $\bM(i, j) = f(\theta_i, \omega_j)$.
We extend the LVM by assuming that $\bW(i, j, l) = f(\theta_i, \omega_j, \psi_l)$, where $f: \Reals^{d_1} \times  \Reals^{d_2} \times  \Reals^{d_3} \to [0, 1]$, $\theta_i \in \Reals^{d_1}$, $\omega_j \in \Reals^{d_2}$ and $\psi_l \in \Reals^{d_3}$. $f$, $\theta$ and $\omega$ are defined identically to that in Section \ref{sec:LVM}. The extension we make to the model now is that $\psi$ is a latent factor that captures the effect of user intent in CTR estimation. 

%

\vspace{2mm}

\noindent \textbf{Algorithm for Tensor Based CTR Estimation.} We know from an extension of the arguments made in \cite{chatterjee2015matrix} that if $\bT$ follows a LVM, then it is ``approximately" low-rank. Thus we postulate that the flattened tensor, i.e. $\bT_{\text{flat}} \in [0, 1]^{m \times (n \times k)}$ is \textit{also ``approximately low-rank"}. Here $\bT_{\text{flat}}$ is the matrix constructed by appending together the $l$ different ``slices" of $\bT$. Thus we propose the following simple algorithm: 

(1) Construct $\bW_{\text{flat}} \in [0, 1]^{m \times (n \times k)}$ from $\bW$ identically to how $\bT_{\text{flat}}$ is defined; (2) Let $\hat{\bT}_{\text{flat}} = \text{ME}(\bW_{\text{flat}})$; (3) Declare $\hat{\bT}(i, j, l) = \hat{\bT}_{\text{flat}}(i, (l -1)m + j)$

\vspace{2mm}

\noindent \textbf{Experimental Verification of Algorithm.} In the Avito dataset, fortunately for our purposes, there is user intent data provided (called ``SearchCategoryID"). There are $40$ user search category IDs present in the dataset. Analogous to Section \ref{sec:avito_dataset}, where we create a $3434 \times 31$ matrix of user locations and ad category, we create an order-3 tensor of dimension $3434 \times 31 \times 40$, by incorporating user intent. We now present two experimental results. 

\begin{figure}[ht]
\centering
\subfigure[]{%
\label{fig:rank_flattened}%
\includegraphics[width=0.9\linewidth,height=4.5cm]{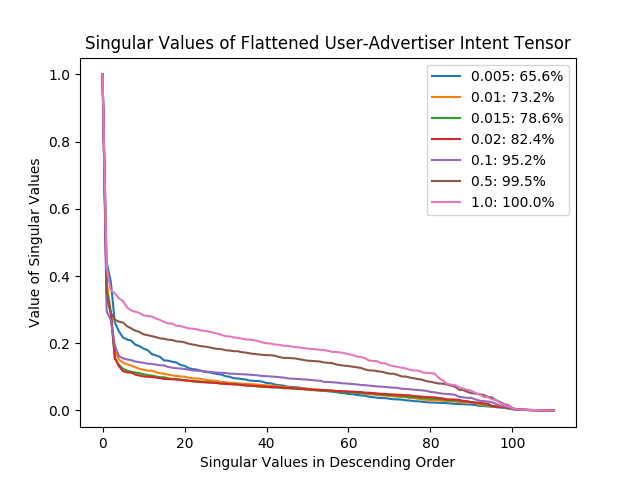}}%
\hfill
\subfigure[]{%
\label{fig:flattened_tensor_vs_matrix_slices}%
\includegraphics[width=0.9\linewidth,height=4.5cm]{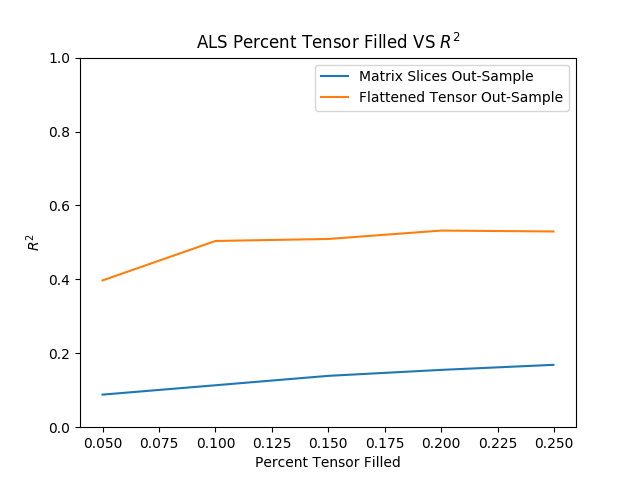}}%
\caption{Figure \ref{fig:rank_flattened} - singular values of flattened tensor for different CTR thresholds \& \% of data retained. Figure \ref{fig:flattened_tensor_vs_matrix_slices} - $R^2$ vs. \% entries filled for our proposed algorithm vs. baseline.}
\end{figure}

In Figure \ref{fig:rank_flattened}, the rank of the flattened user-advertiser-intent matrix (i.e. $\bW_{\text{flat}}$) induced from the tensor (i.e. $\bW$) is approximately 3! This justifies the generalized LVM above. 

In Figure \ref{fig:flattened_tensor_vs_matrix_slices}, we compare our proposed algorithm against a natural baseline; in the baseline, we run matrix estimation (using ALS) for each of the 40 ``slices" separately. By doing so, we do not exploit any of the additional (latent) structure across the user intent dimension. We plot the $R^2$ for the algorithms as a function of the percentage of the tensor that is filled. As desired, we see that the out-of-sample $R^2$ for our algorithm (0.53) is significantly improved over the baseline ($0.17$), indicating that our algorithm does effectively exploit the latent structure across all three dimensions.

%

%
%

\vspace{-5mm}

\section{Conclusion}\label{sec:conclusion}

\vspace{-3mm}

In conclusion, we provide an ``absolute" definition of the VoD, which is independent of a buyer's model for VoD. We operationalize our proposed definition by relying on matrix estimation, and our experiments show that it faithfully (using out-of-sample performance) recovers the VoD. As future work, we intend to open-source and release a Zorro Chrome extension for public use. Lastly, we note that our proposed architecture, where we estimate the VoD in a model agnostic manner, can serve as base to build a general-purpose data recommendation system, i.e., instead of buyers querying Zorro, Zorro recommends data to buyers if estimated VoD is high.

%
%
\bibliographystyle{plain}
\bibliography{usenix2019_v3.1}
%

\end{document}